\documentclass[aps,pre,reprint,longbibliography]{revtex4-1}
\usepackage{graphicx}
\usepackage{amsmath,amssymb}
\usepackage{xcolor}
\usepackage[colorlinks=true,linkcolor=blue]{hyperref}
\usepackage{tikz}
\usetikzlibrary{shapes.geometric, arrows, calc, backgrounds,fit}

\usepackage[utf8]{inputenc}
\usepackage[T1]{fontenc}
\usepackage{mathptmx}

\usepackage{array}
\newcolumntype{C}[1]{>{\centering\arraybackslash}m{#1}}

\begin{document}
\title{Developing a time-domain method for simulating statistical behaviour of many-emitter systems in the presence of electromagnetic field} 
\author{A.R. Hashemi}
\email[]{arezahashemi@gmail.com}
\author{M. Hosseini-Farzad}
\email[]{hosseinif@shirazu.ac.ir}
\affiliation{Department of Physics, College of Sciences, Shiraz University, Shiraz 71946-84795, Iran}

\begin{abstract}
	In this paper, one of the major shortcomings of the conventional numerical approaches is alleviated by introducing the probabilistic nature of molecular transitions into the framework of classical computational electrodynamics. The main aim is to develop a numerical method, which is capable of capturing the statistical attributes caused by the interactions between a group of spontaneous as well as stimulated emitters and the surrounding electromagnetic field. The electromagnetic field is governed by classical Maxwell's equations, while energy is absorbed from and emitted to the (surrounding) field according to the transitions occurring for the emitters, which are governed by time-dependent probability functions. These probabilities are principally consistent with quantum mechanics. In order to validate the proposed method, it is applied to three different test-cases; directionality of fluorescent emission in a corrugated single-hole gold nano-disk, spatial and temporal coherence of fluorescent emission in a hybrid photonic-plasmonic crystal, and stimulated emission of a core-shell SPASER. The results are shown to be closely comparable to the experimental results reported in the literature.
\end{abstract}

\maketitle

\section{Introduction}
Since early developments in the modern optics, light-matter interaction~\cite{FriskKockum2019} has been the central issue in numerous applications, ranging from lasers to photonic computers.
In this field, a vast number of researches has been triggered by Purcell's statement~\cite{Purcell1995,Pelton2015} that the behaviour of a photon emitter is highly sensitive to the surrounding electromagnetic (EM) field; both the decay rate ~\cite{geddes2005radiative} and the emission power~\cite{russell2012large} of the emitter can be effectively controlled by adjusting the EM field.
Such adjustment can be performed using photonic crystal cavities~\cite{PhysRevLett.95.013904,Koenderink:05,doi:10.1021/acs.nanolett.7b05075}, metallic surfaces and nano-particles~\cite{PhysRevLett.89.117401,ribeiro2017artefact}, plasmonic structures~\cite{kinkhabwala2009large,shi2014spatial} and hybrid photonic-plasmonic structures (HPPSs)~\cite{shi2014coherent}.
Another fact to consider is that the emitted light itself can also affect the neighbouring luminescent molecules either directly, as a short-range interaction, or by exciting photonic and/or plasmonic guided modes, which leads to a long range interaction.
In this sense, one needs to deal with complicated time-varying internal and external interactions in a many-body problem~\cite{PhysRevB.91.035306,PhysRevA.4.1791}.
Therefore, further investigations are demanded to develop efficient means to control the response and enhance the emission to a higher degree for applied purposes~\cite{ribeiro2017artefact}.

Considering the practical limitations of the delicate experimental setups, a rather low-cost numerical method can be effectively employed to provide guidelines for experiments~\cite{Bradford:14,Lee:18} while it also advances our fundamental understanding of physical phenomena~\cite{cao2000transition,PhysRevE.98.063304}.
However, the presence of numerous emitters in the vicinity of a nano-structure results in special collective attributes, e.g. coherence, which requires developing an approach to numerically capture the statistical physics of the phenomena.
It is necessary to ensure that the probabilistic nature of molecular transitions is taken into account and consequently, one can utilize a statistical approach to infer the collective attributes. 
In other word, any two emitters of the ensemble that are in the same environmental condition do not deterministically behave in the same way.
In addition, the numerical method should be capable of handling the interaction between emitters and the external EM field as well as the emitter-photon-emitter interactions.

Emitters can be numerically simulated using two different class of methods; those developed based on a macroscopic viewpoint, i.e. using statistically averaged quantities, like effective optical parameters, e.g. permittivity, conductivity, and wave number~\cite{cao2000transition,7769917}, or population densities for molecular energy levels~\cite{Chang:04,662652,PhysRevA.52.3082}. 
In the second class of methods, a microscopic viewpoint is adopted, i.e. one focuses on the behaviour of a single molecule using quantum mechanical methods, like solving the atomic master equation, density function equation, or Schrödinger's equation~\cite{PhysRevB.82.075427,PhysRevA.45.4879,Hong_2015,doi:10.1021/jp1043392}, or modeling the molecular dipole moment of emitters in a classical manner using a damped driven harmonic oscillator differential equation~\cite{PhysRevLett.95.013904,genevet2010large,wang2011optical,bauch2013collective,Hoang2015,javadi2018numerical}. The latter approach has been frequently used in the literature due to its simplicity and capability to be implemented within the framework of conventional numerical methods developed for EM wave propagation, e.g. Finite-Difference Time-Domain (FDTD)~\cite{taflove2005computational}. 

The macroscopic viewpoint is well qualified for simulating a bulk of emitting matters; as an example, Chang and Taflove~\cite{Chang:04} successfully simulated laser gain material by developing a semi-classical approach. Nevertheless, in this viewpoint, the behaviour of numerous microscopic emitters, that are subject to identical conditions, is represented by the averaged macroscopic quantities. Therefore, despite its promising performance for a bulk of active matter, this viewpoint is not a suitable choice for cases of many emitters with individually different dipole orientations and local (especially time-dependent) heterogeneity at the molecular-scale. Moreover, in this class of approaches, the probabilistic nature of problems is lost as a result of the deterministic governing equations. 

On the other hand, numerical methods are developed adopting microscopic viewpoint with the quantum mechanical approach~\cite{doi:10.1021/jp1043392}. 
In these approaches, due to a prohibitive computational cost, the governing equation, e.g. Schrödinger's equation, including the terms corresponding to the external field-emitter and emitter-emitter interactions, can be solved only for a rather small number of emitters.
This shortcoming can be resolved by considering all emitters as identical to each other~\cite{PhysRevB.91.035306}. However, this requires the environmental conditions to be completely the same for all the emitters.
Moreover, acquiring the microscopic viewpoint with a classical approach for modeling the dipoles, one cannot make any distinction between the excitation and emission frequency in cases of three level emitters like fluorescent molecules~\cite{Witthaut_2010,PhysRevLett.106.053601}. 
More importantly, representing the dipole moment of distinct emitters using a single deterministic equation for harmonic oscillator, the statistical nature is lost.

The subject of the present work is to show how the probabilistic nature of the molecular transitions can be microscopically taken into account while the statistical attributes of a rather large set of emitters are macroscopically calculated. 
To this end, a semi-classical approach is proposed, in which transition probabilities are introduced in order to handle the behaviour of emitters, which are considered as single molecules with particular behaviour and distributed in the computational cells. 
To the best of our knowledge, this is the first time that such approach has been acquired. 
It is worth to note that since the presented method is based on generating random samples from probability distributions in a repeated manner, it can be categorized within the broad class of Monte Carlo methods. This class of methods has shown a promising performance in handling the probabilistic nature of many-body problems~\cite{Vishwanath_2002,Lee:18}.
Here, the introduced algorithm is implemented within the framework of the FDTD method and validated against previously reported experimental results for spontaneous and stimulated emissions; directionality and coherence of spontaneous emission and stimulated emission of a SPASER.
It is worth noting that in the present work, the implemented algorithm is developed for a three-level fluorescent molecule as the emitter. 
However, it is straightforward to generalize the algorithm in order to include two- and four-level emitters as well.

\section{Numerical modeling}\label{sec2}
\begin{figure}[!t]
	\centering
	\includegraphics[width=0.9\columnwidth]{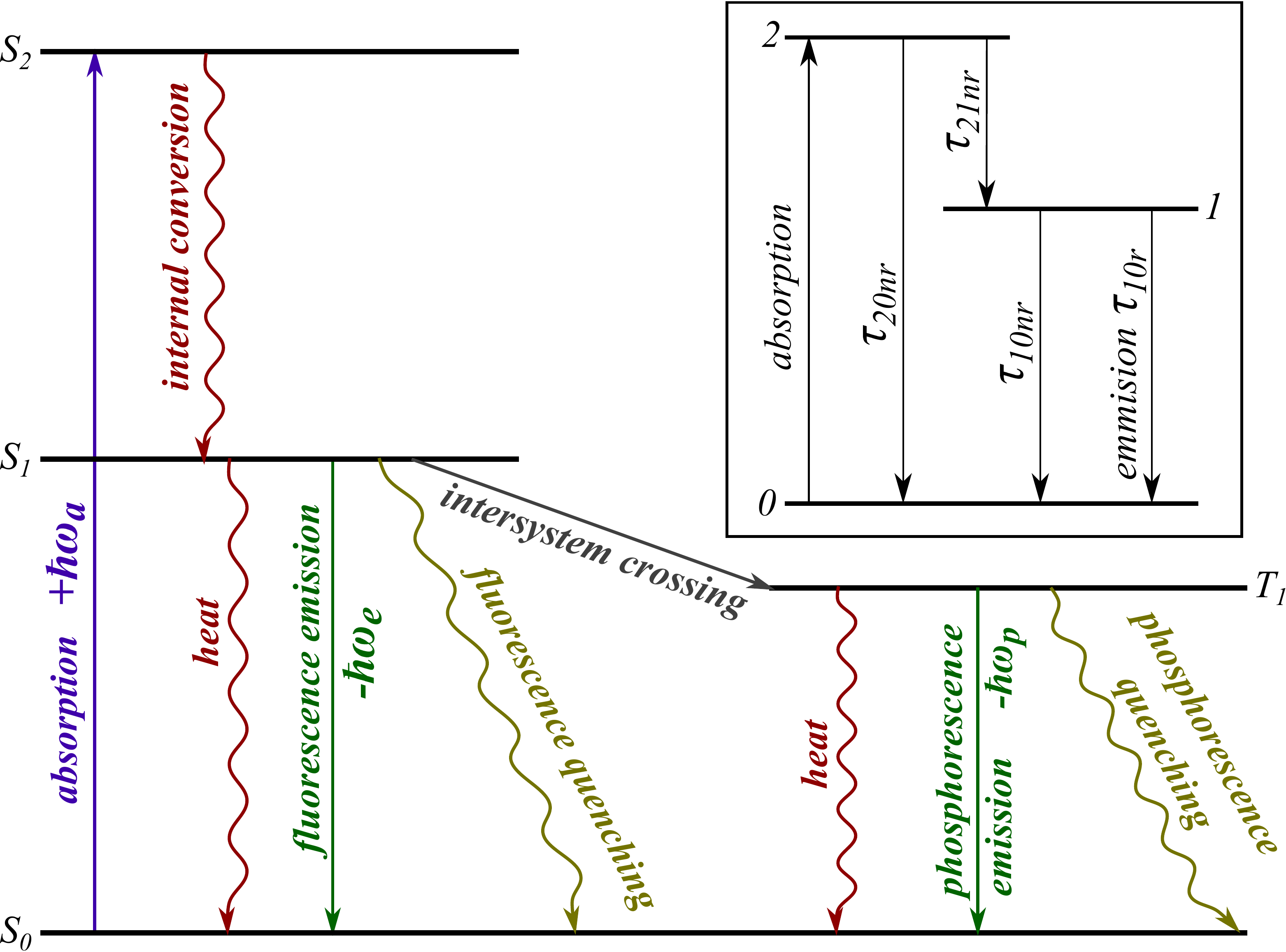}
	\caption{\label{fig1} Jablonski diagram showing electronic transitions of a molecule with radiative and non-radiative transitions from/to singlet ($ S $) and triplet ($ T $) states. The simplified model for a fluorescent molecule is shown in the inset.}
\end{figure}

An emitter with three energy bands, e.g. a fluorescent molecule can be modeled as a three (energy) level system, for which absorption and both radiative and non-radiative transitions occur~\cite{jablonski1933efficiency,albani2008principles} (Fig.~\ref{fig1}). From the quantum mechanical viewpoint, a time dependent probability function can be associated to each of these possible transitions.
In the following, the implementation of these probability functions within the framework of a time-domain method, which is originally developed for simulating EM field propagation, is briefly described for a many-emitter system. 
For more convenience in this paper, the subscripts are corresponding to the energy levels as numbered in Fig.~\ref{fig1}. Moreover, the energy levels of each molecule are identified using three occupation numbers $n_0$, $n_1$, and $n_2$, where at each time-step, only one of them is 1 while the other two are 0. This is due to the fact that at any instance of time, only one of the energy levels can be occupied by the molecule.
Here, $t$ is the total time of the simulation, while $\tilde{t}$ is the time passed from the last transition occurred for the molecule, $P_{ij}$ is the time-dependent probability of transition from level $i$ to level $j$, and the subscripts $r$ and $nr$ refer to the radiative and non-radiative transitions, respectively.

At each time-step, the state is checked for each molecule in the system according to these criteria:
\begin{itemize}
	\item For a molecule in the ground state ($n_0=1$), it is only possible to transit into the second excited state ($0\overset{absorb}{\rightarrow}2$) by absorbing an enough amount of energy. Computationally, this occurs if the randomly generated number $0<r<1$ is less than (or equal to) the corresponding probability $P_{02}(t)$. 
	\item
	For a molecule in the second excited state ($n_2=1$), the non-radiative decay is possible to either the first state ($2\overset{nr}{\rightarrow}1$) or the ground state ($2\overset{nr}{\rightarrow}0$). If random number $r$ is less than (or equal to) $P_{21nr}(\tilde{t})$ transition $2\overset{nr}{\rightarrow}1$ occurs, and on the other hand, the molecule experiences $2\overset{nr}{\rightarrow}0$ if $P_{21nr}(\tilde{t}) < r \leq (P_{21nr}(\tilde{t}) + P_{20nr}(\tilde{t}))$.
	\item 
	For a molecule in the first excited state ($n_1=1$), both the radiative and non-radiative decays to the ground state are possible. Transition $1\overset{nr}{\rightarrow}0$ occurs if $r \leq P_{10nr}(\tilde{t})$. Else, if $P_{10nr}(\tilde{t}) < r \leq (P_{10nr}(\tilde{t}) + P_{10r}(\tilde{t}))$, the radiative transition of $1\overset{r}{\rightarrow}0$ takes place and consequently, a wave packet with the central frequency equal to emission frequency of the fluorophore $\omega_e$ is emitted from a point source at the position of the emitting molecule.
\end{itemize}
Once a transition occurs, the occupation numbers $n_0$, $n_1$, and $n_2$ are correspondingly reset, e.g., $0\overset{absorb}{\rightarrow}2$ is associated with resetting numbers as $n_0=0$ and $n_2=1$.
 During emission, the re-excitation of the corresponding molecule is prevented by setting its emission flag to \emph{on}, which means the state of the molecule is ignored and the emission continues until the emitted wave packet vanishes. At this moment, the emission flag is off and the molecule is at its ground state.

The probabilities associated with the above-mentioned transitions between these levels are estimated as described in the following. It is evident that these probabilities can be modified into more sophisticated functions that are obtained by quantum mechanical analysis of molecules. However, the following formulas show the simplest probability functions that satisfy the physical requirements and work successfully for the test-cases solved in this work.

\begin{figure}[!t]
	\centering
	\includegraphics[width=0.95\columnwidth]{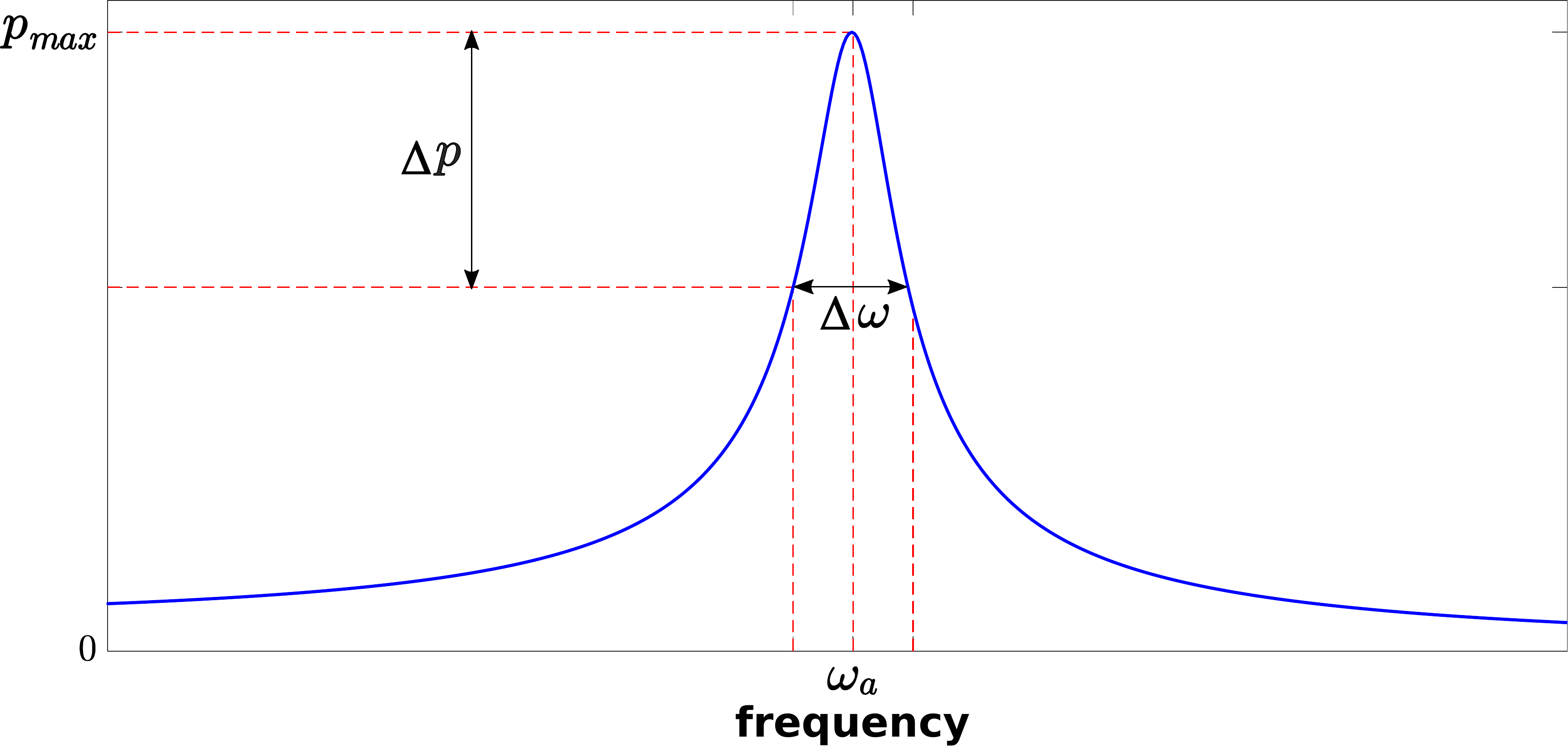}
	\caption{\label{fig:pomega}Amplitude of $\mathbf{p}$ as a function of $\omega$.}
\end{figure}

\textit{Absorption}:
For a molecule with the frequency of the maximum absorption, $\omega_a$, the absorption of energy from the field and excitation of the molecule becomes more probable if the electric field imposed at the position of molecule, $\mathbf{E}(t)$ incorporates a frequency component tending to $\omega_a$.
In order to examine this possibility, the occurrence of the resonance of a harmonic oscillator driven by $\mathbf{E}(t)$ is checked. This oscillator resembles the electric dipole moment of the molecule, $\mathbf{p}$. Within the context of the time-domain method used in the present work, i.e. FDTD, the equation that governs the response of the harmonic oscillator is considered as an auxiliary differential equation (ADE)~\cite{kashiwa1990treatment}. It must be highlighted that using the aforementioned ADE is the most efficient way to check the frequency components of $\mathbf{E}(t)$ against $\omega_a$. The implemented ADE is 
\begin{equation}
\ddot{{p}}+\gamma\dot{{p}}+\omega_a^2{p}=({e^2}/{m})\mathbf{E}(t)\cdot\hat{\mathbf{p}}.\label{eq1}
\end{equation}
Here, $\gamma$ is a damping factor and $e$ and $m$ are the charge and mass of electron, respectively.
Vector $\hat{\mathbf{p}}$ is a unit vector representing the orientation of dipole moment of the emitter, i.e. $\mathbf{p}=p\hat{\mathbf{p}}$. 
The magnitude of the dipole moment is initially zero and updated (at time step $n+1$) using a second-order central time-marching scheme as done for Maxwell's equations in the adopted FDTD method~\cite{taflove2005computational}. 
At the onset of resonance, the amplitude of $\mathbf{p}$ tends to its maximum value $p_{max}$ and consequently, the transition from level 0 to 2 becomes more probable. Therefore, the corresponding probability is estimated as
\begin{equation}
P_{02}(t_n)=\exp\left(-{\left({p}(t_n)-p_{max}\right) ^2}/{2\sigma^2}\right),\label{eq3}
\end{equation}
where $t_n$ represents time at $n$th time-step, i.e $t_n=n\Delta t$, and $\sigma$ is determined in terms of the absorption bandwidth of the fluorophore, $\Delta\omega_a$. 
In order to derive an equation for $\sigma$, one can consider the external field in its simplest form $\mathbf{E}(t)=\mathbf{E_0}\sin(\omega t)$ and analytically solve Eq.~(\ref{eq1}) for the amplitude of $\mathbf{p}$, which depends on $\omega$ as
\begin{equation}
{p}(\omega)=\frac{({e^2}/{m})|E_0|}{\sqrt{\left(\omega_a^2-\omega^2\right)^2+\left(\gamma\omega\right)^2}}.\label{eq3_1}
\end{equation}
As illustrated in Fig.~(\ref{fig:pomega}), for an absorption bandwidth of $\Delta\omega_a$, an amplitude interval of $\Delta p$ is defined as
\begin{equation}
\Delta p=p_{max}-{p}(\omega_a-\tfrac{\Delta\omega_a}{2}).
\end{equation}
In this way, the standard deviation $\sigma$ is considered equal to $\Delta p$.
It must be noted that since $p_{max}\propto \frac{1}{\gamma\omega_a}$, both $p_{max}$ and damping factor $\gamma$ are arbitrary factors which should be determined correspondingly.

Upon the occurrence of resonance, energy is also absorbed from the external EM field, which is governed by Ampere-Maxwell's relation
$$\varepsilon\dot{\mathbf{E}}=\nabla\times\mathbf{H}-\sum_{i=1}^{N}\dot{\mathbf{p}}_i.$$
Here, $N$ denotes the number of fluorophores associated with the computational grid cell, which is proportional to the density of the fluorescent material at the same point within the FDTD discretized domain. It is possible to have various fluorescent densities at different locations of the structure.
If referring to Eq.~\ref{eq3} transition $0\overset{absorb}{\rightarrow}2$ occurs, $p$ and its first temporal derivative are set to zero.
On the other hand, if $0\overset{absorb}{\rightarrow}2$ does not occur, the dipole moment of the molecule is updated via equation~\ref{eq1}.

\textit{Non-radiative transitions}:
For the present model, three non-radiative transitions are considered; $2\overset{nr}{\rightarrow}1$, $2\overset{nr}{\rightarrow}0$ and $1\overset{nr}{\rightarrow}0$. The corresponding probabilities are estimated as
\begin{eqnarray}
P_{21nr}(\tilde{t}) = &A\left(1-\exp\left(-\frac{\tilde{t}}{\tau_{21nr}}\right)\right),\label{eq4}\\
P_{20nr}(\tilde{t}) = &B\left(1-\exp\left(-\frac{\tilde{t}}{\tau_{20nr}}\right)\right),\label{eq5}\\
P_{10nr}(\tilde{t}) = &C\left(1-\exp\left(-\frac{\tilde{t}}{\tau_{10nr}}\right)\right).\label{eq6}
\end{eqnarray}
Here, $\tau_{ijnr}$ represents the time-constant for a non-radiative decay between levels $ i $ and $ j $. The asymptotic behavior of these functions guaranties a definite decay at an infinitely long time. 

\textit{Radiative transition}:
Radiative transition is only considered as a decay from level 1 to level 0, for which the corresponding probability is estimated as
\begin{eqnarray}
P_{10r}(\tilde{t}) = &D\left(1-\exp\left(-\frac{\tilde{t}}{\tau_{10r}}\right)\right),\label{eq7}
\end{eqnarray}
where $\tau_{10r}$ is the time constant of the radiative decay.
Once the transition occurs, a wave-packet is emitted with a central-frequency equal to the emission frequency of the fluorophore, $\omega_e$. This is implemented as a soft point source, which is a vector function, $\mathbf{F}(\tilde{t})$, that is superimposed to the electric (and/or magnetic) field at the position of the molecule. This function is formulated as
\begin{eqnarray}
\mathbf{F}(\tilde{t})=\mathbf{F}_0\exp\left(-\frac{(\tilde{t}-t_0)^2}{\sigma_e^2}\right)\sin(\omega_e(\tilde{t}-t_0)),\label{eq:emPulse}
\end{eqnarray}
where $\mathbf{F}_0=F_0\hat{\mathbf{n}}$. In this equation, $\hat{\mathbf{n}}$ is a randomly oriented unit vector and $F_0$ is determined in a manner that the total energy of the wave-packet is equal to the energy of an emitted photon. Here, $\sigma_e$ is associated with the band-width of the emission and $t_0\approx 3\sigma_e$ is a time-offset, which guarantees that the at $\tilde{t}=0$, $\mathbf{F}$ approaches zero.

The normalization factors $A,\, B,\, C,\, D$ are calculated based on two physical concepts; first, it is impossible for a molecule to permanently stay at an excited state and consequently, $A+B=1$ and $C+D=1$. Second, the probability ratio of the transitions is inversely related to the corresponding decay times, $\tau_{ij}$. Therefore, one can obtain
\begin{eqnarray}
&A={\tau_{21nr}}/({\tau_{21nr}+\tau_{20nr}}),\quad &B={\tau_{20nr}}/({\tau_{21nr}+\tau_{20nr}}),\nonumber\\
&C={\tau_{10nr}}/({\tau_{10nr}+\tau_{10r}})\quad \text{and}\quad &D={\tau_{10r}}({\tau_{10nr}+\tau_{10r}}).\nonumber
\end{eqnarray}
The flowcharts illustrated in Figures~\ref{fig:Flow1} and \ref{fig:Flow2}, present the implemented algorithm in more details.

\begin{figure}[!t]
	\centering
	\begin{tikzpicture}[node distance=7ex,thick,scale=0.8, every node/.style={scale=0.8}]
	\tikzstyle{startstop} = [rectangle, rounded corners=1.2ex, minimum width=10ex, minimum height=2ex,text centered, draw=black, fill=red!30]
	\tikzstyle{io} = [trapezium, trapezium left angle=70, trapezium right angle=110, minimum width=10ex, minimum height=2ex, text centered, draw=black, fill=blue!30]
	\tikzstyle{process} = [rectangle, minimum width=10ex, minimum height=2ex, text centered, draw=black, fill=orange!30]
	\tikzstyle{decision} = [diamond, minimum width=10ex, minimum height=2ex, aspect=2, text centered, draw=black, fill=green!30, inner sep=1pt]
	\tikzstyle{conect} = [circle, minimum width=0.01ex, minimum height=0.01ex, text width=0.01ex, draw=black, fill=yellow!80!black]
	\tikzstyle{con} = [circle, minimum width=1pt, minimum height=1pt, text width=1pt, draw=black, fill=yellow]
	\tikzstyle{arrow} = [thick,->,>=stealth]%
	\tikzstyle{line} = [thick,-]
	\node (start) [startstop] {Start};	
	\node (pre) [process, below of=start, text width=20ex] {FDTD preprosses and fields initializations};
	\draw [arrow] (start) -- (pre);
	\node (init0) [process, below of=pre,yshift=-1ex] {$i=0$};
	\draw [arrow] (pre) -- (init0);
	\node (init1) [process, below of=init0, text width=22ex] {initialize populations\\ $n_{0i}=1,n_{1i}=0,n_{2i}=0$};
	\draw [arrow] (init0) -- (init1);
	\node (init) [process, text width=22ex, below of=init1,yshift=-1.5ex] {initialize populations\\ $\mathbf{p}_i=0,\mathbf{\dot{p}}_i=0$};
	\draw [arrow] (init1) -- (init);
	\node (initd) [decision, below of=init,yshift=-6ex, inner sep=0pt, text width=11ex] {{\small $i\overset{?}{<}$ number of emitters}};
	\draw [arrow] (init) -- (initd);
	\node (init2) [process, above of=initd, xshift=-15ex, yshift=-1ex] {$i=i+1$};
	\draw [arrow] (initd) -| node[below,xshift=2ex]{yes} (init2);
	\draw [arrow] (init2) |- (init1);
	\tikzset{blue dotted/.style={draw=blue!50!white, line width=1pt, dashed, inner sep=6ex, rectangle, rounded corners}};
	\node (first dotted box) [blue dotted, fit = (init0) (init1) (init) (initd) (init2)] {};
	\node at (first dotted box.west) [above, inner sep=1ex, rotate=90] {{emitters initialization}};
	\node (pro0) [process, below of=initd, yshift=-6ex] {$t=0$};
	\draw [arrow] (initd) -- node[right,yshift=1.2ex]{no} (pro0);
	\node (pro1) [process, below of=pro0, yshift=-1.5ex, text width=20ex, inner xsep=1pt] {FDTD updates for EM fields, metals, PMLs, sources, ...};
	\draw [arrow] (pro0) -- (pro1);
	\node (pro2) [blue dotted, draw=blue!90!white, below of=pro1, yshift=-6ex, text width=22ex, inner sep=5pt] {\centering{\textbf{emitters state-updating} \\(depicted in Fig.~\ref{fig:Flow2})}};
	\draw [arrow] (pro1) -- (pro2);
	\node (mtl) [decision, below of=pro2, yshift=-8ex, inner sep=1pt] {{\small $t\overset{?}{\leqslant}$ final time step}};
	\draw [arrow] (pro2) -- (mtl);
	\node (pro3) [process, above of=mtl, xshift=-15ex, yshift=-1ex] {$t=t+\delta t$};
	\draw [arrow] (mtl) -| node[below,xshift=2ex]{yes} (pro3);
	\draw [arrow] (pro3) |- (pro1);
	\node (second dotted box) [blue dotted, fit = (pro0) (pro1) (pro2) (pro3) (mtl)] {};
	\node at (second dotted box.west) [above, inner sep=1ex, rotate=90] {{FDTD time-marching loop}};
	\node (post) [process, below of=mtl, yshift =-5ex, text width=20ex] {FDTD postprossing};
	\draw [arrow] (mtl) -- node[right,yshift=1.0ex]{no} (post);
	\node (out1) [io, below of=post, yshift=.5ex] {Output};
	\draw [arrow] (post) -- (out1);
	\node (stop) [startstop, below of=out1, yshift=.5ex] {Stop};
	\draw [arrow] (out1) -- (stop);	
	
	\end{tikzpicture}
	\caption{\label{fig:Flow1} The modified FDTD algorithm for implementing the procedure required for simulating emitters-EM field interactions. All emitters are at the ground-state at the beginning of simulation and updated during the main time-marching loop. The state-updating steps are elaborated in Fig.~\ref{fig:Flow2}.}
\end{figure}
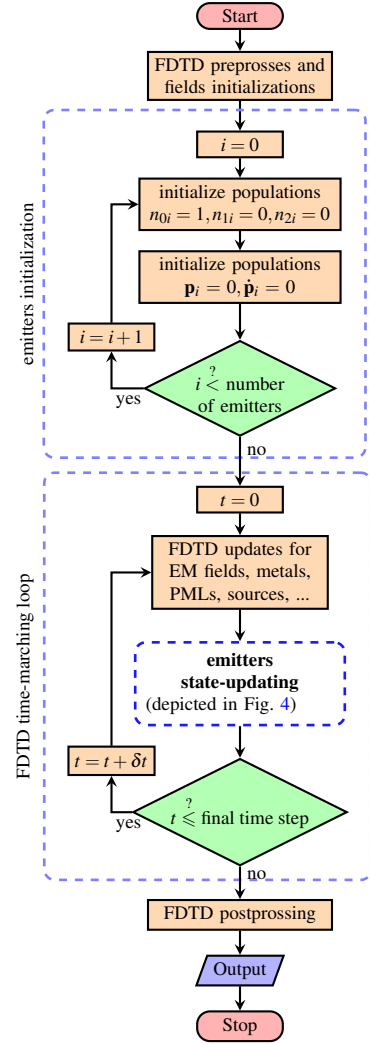

\begin{figure*}[!t]
	\centering
	\begin{tikzpicture}[node distance=7ex,thick,scale=0.8, every node/.style={scale=0.8}]
	\tikzstyle{startstop} = [rectangle, rounded corners=1.2ex, minimum width=10ex, minimum height=2ex,text centered, draw=black, fill=red!30]
	\tikzstyle{io} = [trapezium, trapezium left angle=70, trapezium right angle=110, minimum width=10ex, minimum height=2ex, text centered, draw=black, fill=blue!30]
	\tikzstyle{process} = [rectangle, minimum width=10ex, minimum height=2ex, text centered, draw=black, fill=orange!30]
	\tikzstyle{decision} = [diamond, minimum width=10ex, minimum height=2ex, aspect=2, text centered, draw=black, fill=green!30, inner sep=1pt]
	\tikzstyle{conect} = [circle, minimum width=0.001ex, minimum height=0.001ex, text width=0.01ex, inner sep=2pt, draw=black, fill=yellow]
	\tikzstyle{con} = [circle, minimum width=1pt, minimum height=1pt, text width=1pt, draw=black, fill=yellow!80!black]
	\tikzstyle{arrow} = [thick,->,>=stealth]%
	\tikzstyle{line} = [thick,-]
	
	\node (pro0) [process] {$i=0$};
	\draw [arrow] ++(0,6.5ex)-- (pro0);
	\node (emdes) [decision, below of=pro0, yshift=-7ex, inner sep=0pt, text width=14ex, align=center] {$EF_i$\\{\small (emission flag of the $i$th emitter)}\\is on?};
	\draw [arrow] (pro0) -- (emdes);
	\node (emup) [process, right of=emdes, xshift=25ex, text width=19ex, inner xsep=1pt] {$\mathbf{E}_i(t)=\mathbf{E}_i(t)+\mathbf{F}_i(\tilde{t}_i)$\\$\tilde{t}_i=\tilde{t}_i+1$};
	\draw [arrow] (emdes) -- node[above] {yes} (emup);
	\node (findes) [decision, right of=emup, xshift=18ex]{$\tilde{t}_i\overset{?}{\geqslant}$ emission time};
	\draw [arrow] (emup) -- (findes);
	\node (flagoff) [process, right of=findes, xshift=15ex] {$EF_i=$ off};
	\draw [arrow] (findes) -- node[above] {yes} (flagoff);
	\node (con1) [conect, below of=flagoff] {};
	\draw [arrow] (flagoff) -- (con1);
	\draw [arrow] (findes) |- node[right, yshift=1ex, xshift=2ex] {no} (con1);
	\node (randgen) [process, below of=emdes, yshift=-10ex, text width=15ex, inner xsep=1pt] {{\small generate a random number}\\$0<r_i\leq 1$};
	\draw [arrow] (emdes) -- node[right] {no} (randgen);
	\node (n0des) [decision, right of=randgen, xshift=11ex] {$n_{0i}\overset{?}{=}1$};
	\draw [arrow] (randgen) -- (n0des);
	\node (p02des) [decision, right of=n0des, xshift=16ex, text width=11ex, inner sep=0pt] {$r_i\overset{?}{<}P_{02i}(t)$ {\small (Eq.~\ref{eq3})}};
	\draw [arrow] (n0des) -- node[above] {yes} (p02des);
	\node (n0pro1) [process, right of=p02des, xshift=16ex, text width=13ex, inner xsep=1pt] {$n_0=0,n_2=1$ $\mathbf{p}_i=0,\mathbf{\dot{p}}_i=0$ $\tilde{t}_i=0$};
	\draw [arrow] (p02des) -- node[above] {yes} (n0pro1);
	\node (n0pro2) [process, below of=n0pro1, yshift=-1ex] {update $\mathbf{p}_i$ {\small (Eq.~\ref{eq1})}};
	\draw [arrow] (p02des) |- node[right, yshift=1ex] {no} (n0pro2);
	\node (con2) [conect, below of=con1, yshift=-3ex] {};
	\draw [arrow] (n0pro1) -- (con2);
	\draw [arrow] (con1) -- (con2);
	\node (con3) [conect, below of=con2,yshift=-1ex] {};
	\draw [arrow] (n0pro2) -- (con3);
	\draw [arrow] (con2) -- (con3);
	\node (n1des) [decision, below of=n0des, yshift=-10ex] {$n_{1i}\overset{?}{=}1$};
	\draw [arrow] (n0des) -- node[left] {no} (n1des);
	\node (p10des1) [decision, below of=p02des, yshift=-10ex, text width=12ex, inner sep=0pt] {$r_i\overset{?}{<}P_{10nri}(\tilde{t}_i)$ {\small (Eq.~\ref{eq6})}};
	\draw [arrow] (n1des) -- node[above] {yes} (p10des1);
	\node (n1pro1) [process, right of=p10des1, xshift=17ex, text width=13ex] {$n_1=0,n_0=1$ $\tilde{t}_i=0$};
	\draw [arrow] (p10des1) -- node[above,xshift=-.5ex] {yes} (n1pro1);
	\node (con4) [conect, below of=con3,yshift=-2ex] {};
	\draw [arrow] (n1pro1) -- (con4);
	\draw [arrow] (con3) -- (con4);
	\node (p10des2) [decision, below of=p10des1, yshift=-11ex, text width=23ex, inner sep=0pt] {$r_i\overset{?}{<}(P_{10nri}(\tilde{t}_i)+P_{10ri}(\tilde{t}_i))$\\{\small (Eqs.~\ref{eq6} \& \ref{eq7})}};
	\draw [arrow] (p10des1) -- node[right] {no} (p10des2);
	\node (n1pro2) [process, right of=p10des2, xshift=19ex, text width=9ex] {$n_1=0$ $n_0=1$ $\tilde{t}_i=0$ $EF_i=$ on};
	\draw [arrow] (p10des2) -- node[above,xshift=-.5ex] {yes} (n1pro2);
	\node (con5) [conect, below of=con4,yshift=-11ex] {};
	\draw [arrow] (n1pro2) -- (con5);
	\draw [arrow] (con4) -- (con5);
	\node (n1pro3) [process, below of=n1pro2, yshift=-3ex, text width=9ex] {$\tilde{t}_i=\tilde{t}_i+1$};
	\draw [arrow] (p10des2) |- node[right, yshift=1ex, xshift=1ex] {no} (n1pro3);
	\node (con6) [conect, below of=con5, yshift=-3ex] {};
	\draw [arrow] (n1pro3) -- (con6);
	\draw [arrow] (con5) -- (con6);
	\node (n2des) [decision, below of=n1des, yshift=-29ex] {$n_{2i}\overset{?}{=}1$};
	\draw [arrow] (n1des) -- node[left] {no} (n2des);
	\node (p20des) [decision, below of=p10des2, yshift=-11ex, text width=12ex, inner sep=0pt] {$r_i\overset{?}{<}P_{20nri}(\tilde{t}_i)$ {\small (Eq.~\ref{eq5})}};
	\draw [arrow] (n2des) -- node[above] {yes} (p20des);
	\node (n2pro1) [process, right of=p20des, xshift=17ex, text width=13ex] {$n_2=0,n_0=1$ $\tilde{t}_i=0$};
	\draw [arrow] (p20des) -- node[above,xshift=-.5ex] {yes} (n2pro1);
	\node (con7) [conect, below of=con6,yshift=-1ex] {};
	\draw [arrow] (n2pro1) -- (con7);
	\draw [arrow] (con6) -- (con7);
	\node (p21des) [decision, below of=p20des, yshift=-11ex, text width=23ex, inner sep=0pt] {$r_i\overset{?}{<}(P_{21ri}(\tilde{t}_i)+P_{20nri}(\tilde{t}_i))$\\{\small (Eqs.~\ref{eq4} \& \ref{eq5})}};
	\draw [arrow] (p20des) -- node[right] {no} (p21des);
	\node (n2pro2) [process, right of=p21des, xshift=19.5ex, text width=9ex] {$n_2=0$ $n_1=1$ $\tilde{t}_i=0$};
	\draw [arrow] (p21des) -- node[above,xshift=-.5ex] {yes} (n2pro2);
	\node (con8) [conect, below of=con7,yshift=-11ex] {};
	\draw [arrow] (n2pro2) -- (con8);
	\draw [arrow] (con7) -- (con8);
	\node (n2pro3) [process, below of=n2pro2, yshift=-3ex, text width=9ex] {$\tilde{t}_i=\tilde{t}_i+1$};
	\draw [arrow] (p21des) |- node[right, yshift=1ex, xshift=1ex] {no} (n2pro3);
	\node (con9) [conect, below of=con8, yshift=-3ex] {};
	\draw [arrow] (n2pro3) -- (con9);
	\draw [arrow] (con8) -- (con9);
	\node (fludes) [decision, below of=randgen, yshift=-70ex, inner sep=1pt] {$i\overset{?}{<}$ number of emitters};
	\draw [arrow] (con9) |- ++(-53ex,-4ex) |- (fludes);
	\node (ipro) [process, above of=fludes, xshift=-22ex, yshift=5ex] {$i=i+1$};
	\draw [arrow] (fludes) -| node[below,xshift=6ex]{yes} (ipro);
	\draw [arrow] (ipro) |- (emdes);
	\draw [arrow] (fludes) -- node[right,yshift=1.2ex]{no} ++(0,-13ex);
	\tikzset{blue dotted/.style={draw=blue!50!white, line width=1pt, dashed, inner sep=14ex, rectangle, rounded corners}};
	\node (first dotted box) [blue dotted, fit=(pro0) (ipro) (fludes) (emdes) (flagoff)] {};
	\node at (first dotted box.west) [above, inner sep=1ex, rotate=90] {emitters state-updating};
	\end{tikzpicture}
	\caption{\label{fig:Flow2} State-updating procedure for emitters. In this flowchart, $EF_i$ is the emission flag of $i$th emitter, which is \emph{on} while the wave-packet is emitting. $\mathbf{E}_i(t)$ is the electric field at the position of the $i$th emitter, $\mathbf{F}_i(\tilde{t}_i)$ is the emitted electric wave function, and $t_i$ is the time elapsed since the $i$th emitter has been transited to its current state. In each time step, if the emission flag of an emitter (e.g. the $i$th one) is on, the emission continues. If the flag is off, the state of the emitter should be checked. In case the emitter is at the ground-state, the absorption probability $P_{02i}(t)$ is compared to a randomly generated number, $r_i$, and consequently, either the transition to level 2 takes place or the dipole moment of the emitter is updated. In other cases (i.e. the emitter is at level 1 or 2) the same procedure is followed with the corresponding probability function.}
\end{figure*}
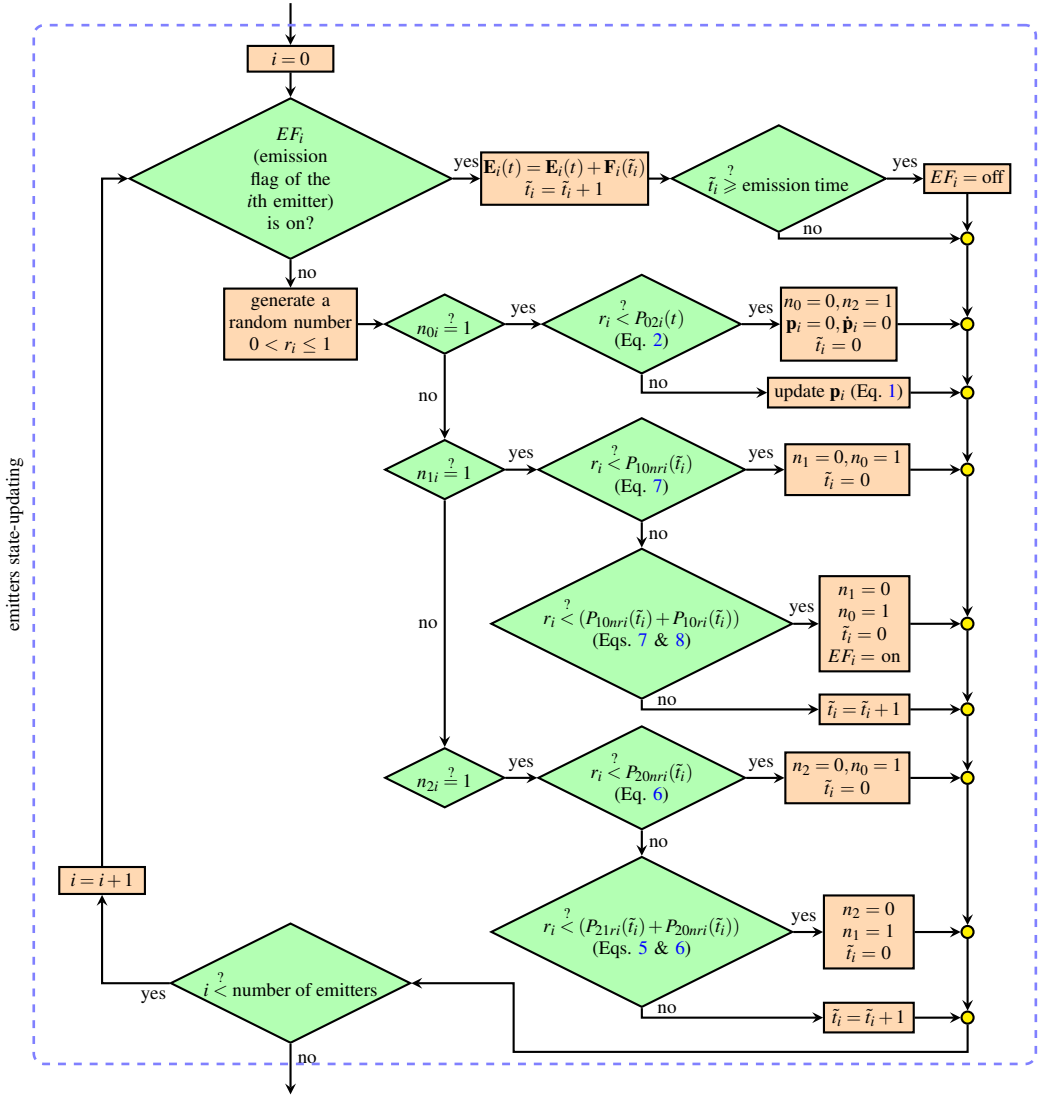

In this work, the propagation of EM fields in a three-dimensional domain and successive time-steps is simulated using an FDTD package, which is developed in C++ language. Convolutional perfectly matched layer (CPML)~\cite{roden2000convolutional,berenger2002application} boundary condition is used for reducing the reflection from exterior boundaries of the simulation domain.
Whenever needed, the total-field/scattered-field technique and the Drude-Lorentz model are implemented to handle the incident field and dispersive materials, respectively~\cite{taflove2005computational}.
In all simulations, the electric and magnetic fields are initially set to zero.
It is worth noting that the implementation of the same algorithm is also possible for other numerical methods, e.g. the finite-element time-domain method.

\section{Applications for many spontaneous emitters}
In the following, the proposed method is applied to two phenomena observed for many-emitter systems; directionality of fluorescence in the presence of a plasmonic nano-structure~\cite{aouani2011plasmonic} and fluorescence coherence obtained by utilizing an HPPS~\cite{shi2014coherent}. The present method is validated by comparing the numerical results with the corresponding experimental results reported in the literature.

Here, the spatial discretization of the solution domain is set considering the criteria for minimizing the dispersion error of the FDTD method~\cite{taflove2005computational} while resolving all structural details. Governing equations are solved for a three-dimensional Cartesian mesh with Yee cells of $\Delta x=\Delta y=\Delta z$. The time-step is set according to the Courant condition. The physical properties of the specific emitters simulated in the present work are shown in Table~\ref{tab:flupara}.
\begin{table*}
	\begin{ruledtabular}
		\begin{tabular}{l|C{10ex}C{10ex}C{10ex}C{12ex}C{8ex}C{8ex}C{8ex}C{8ex}}
			&excitation wavelength&emission wavelength&quantum yield&excited state life-time&$\tau_{10r}$&$\tau_{10nr}$&$\tau_{21nr}$&$\tau_{20nr}$\\
			\colrule
			Alexa Fluor 647&650~nm&672~nm&0.3&1.04~ns&3.5~ns&1.7~ns&0.01~ns&11.4~ns\\
			Rhodamine 6G&525~nm&550~nm&0.95&4.08~ns&4.3~ns&92.7~ns&0.01~ns&824.1~ns\\
			Sulfohodamine 101&575~nm&591~nm&0.8&4.2~ns&5.2~ns&23.8~ns&0.01~ns&257.5~ns\\
			Oregon Green 488&494~nm&524~nm&0.9&4.3~ns&4.8~ns&48.8~ns&0.01~ns&265.9~ns
		\end{tabular}
	\end{ruledtabular}
	\caption{\label{tab:flupara}Optical properties of emitters used in test-cases.}
\end{table*}

\subsection{Directional spontaneous emission}
\begin{figure}[!t]
	\centering
	\includegraphics[width=0.9\columnwidth]{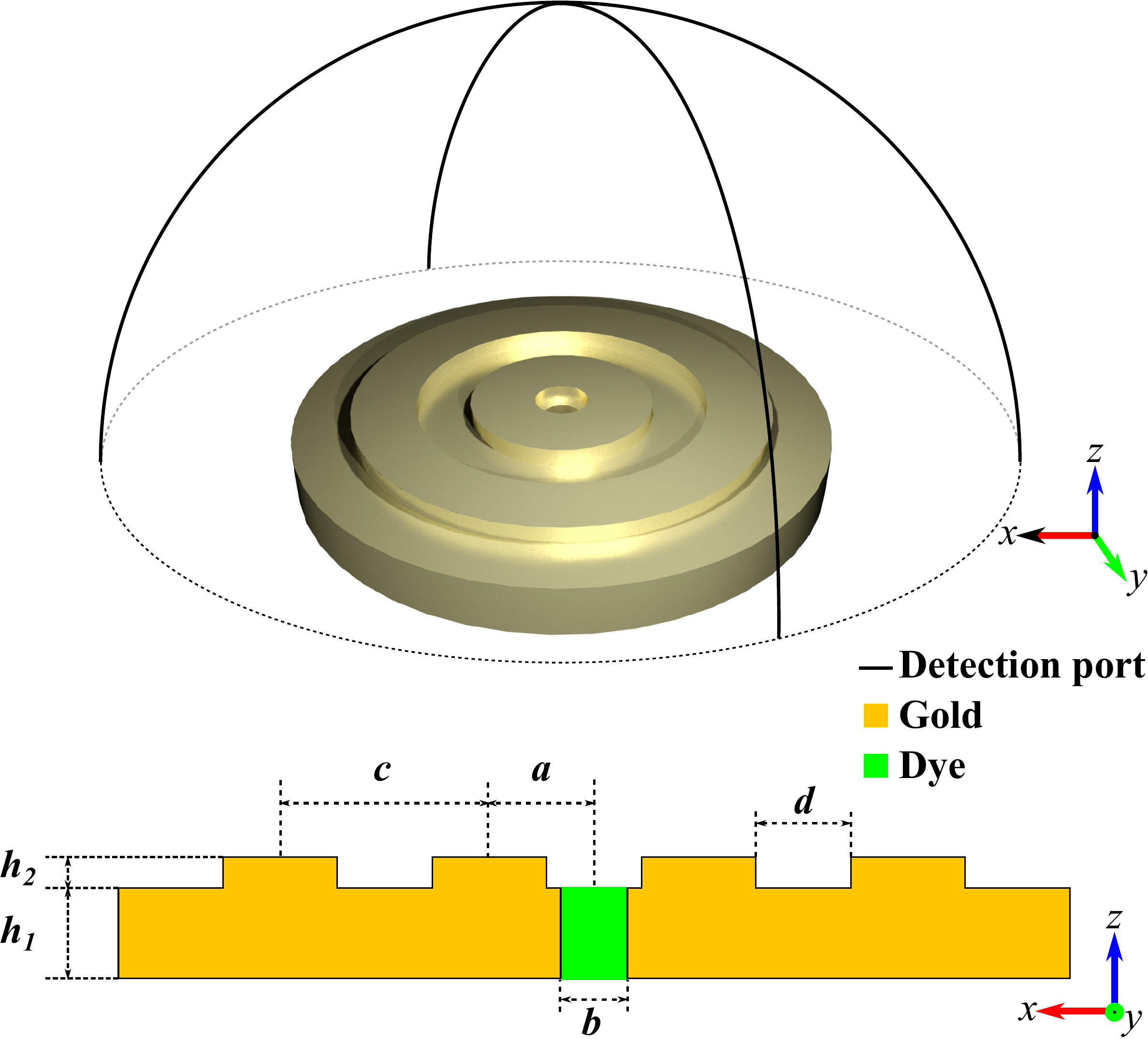}
	\caption{\label{fig:heykel}Schematic of the plasmonic nano-disk with the fluorescent molecules positioned at the central hole of it as proposed by Aouani et al.~\cite{aouani2011plasmonic}. The three-dimensional view is shown on top and the cross-sectional view is presented at the bottom. The geometrical parameters are set according to~\cite{aouani2011plasmonic} as $a=220$~nm, $b=140$~nm, $c=440$~nm, $d=200$~nm, $h_1=190$~nm and $h_2=65$~nm.
	}
\end{figure}

In this section, the central hole of a gold nano-disk with two concentric grooves is filled by excited fluorescent molecules as shown in Fig.~\ref{fig:heykel}. The capability of this plasmonic system in producing directional emission has been experimentally studied by Aouani et al.~\cite{aouani2011plasmonic}. The geometrical parameters (see Fig.~\ref{fig:heykel}), as well as the physical properties of the base structure (gold) and the fluorescent (Alexa Fluor 647 and Rhodamine 6G) molecules, are set the same as those reported in~\cite{aouani2011plasmonic}. 
	The properties of Alexa Fluor 647 and Rhodamine 6G are set according to~\cite{doi:10.1021/bc025600x} and~\cite{doi:10.1562/0031-8655(2002)0750327FQYATR2.0.CO2}, respectively (see Table~\ref{tab:flupara}).
	In this test-case, the caution is that the resulted emission distribution is hardly distinguishable due to the masking effect of the incident pump beam.
	Here, in order to eliminate the need for a post-processing procedure, the fluorophore molecules are considered to be initially-excited ($n_2=1$ for all molecules). In this way, there is no need to impose a pump beam.

\begin{figure}[!t]
	\centering
	\includegraphics[width=0.99\columnwidth]{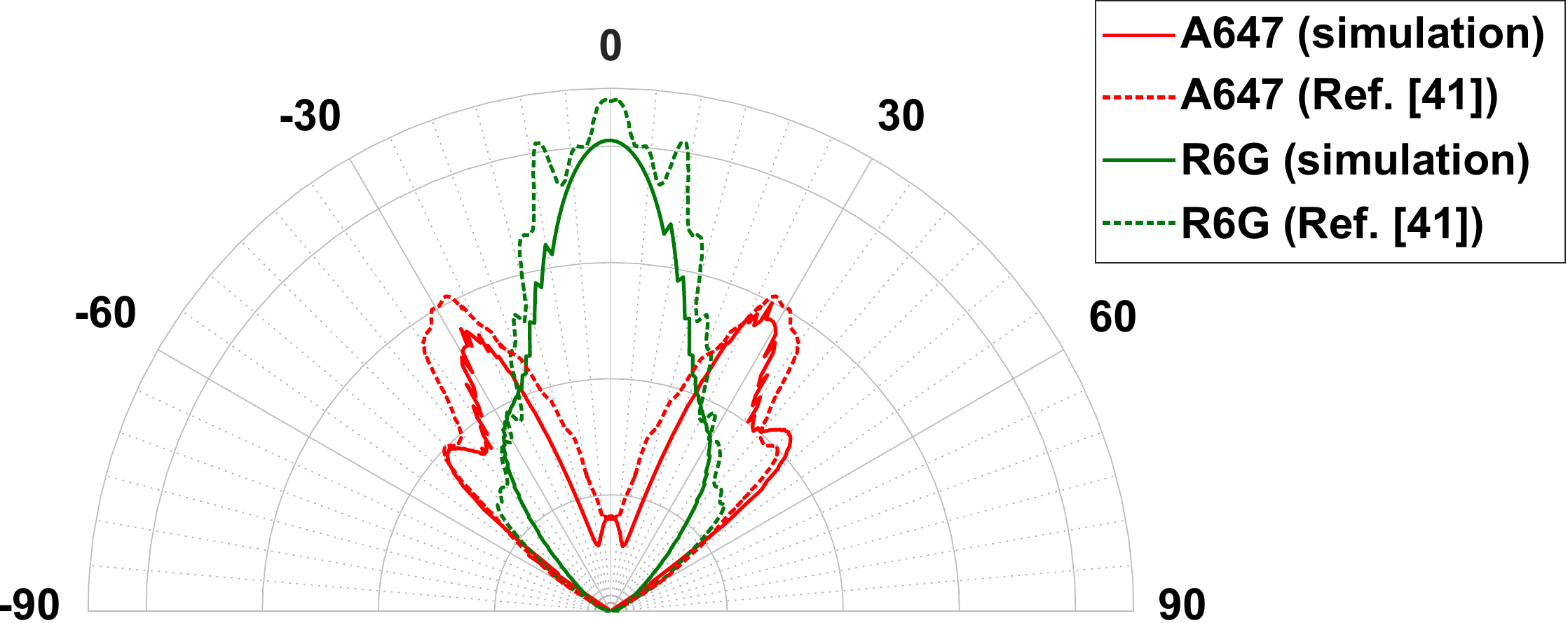}
	\caption{\label{fig:heykel_emi}Polar distribution of the long-time and ensemble averaged intensity for emissions detected from the plasmonic nano-disk. Red and green curves correspond to the results obtained for Alexa Fluor 647 ($\lambda_{em} = 670$~nm) and Rhodamine 6G ($\lambda_{em} = 560$~nm), respectively. Numerical results (solid-lines) are compared with the results reported by Aouani et al.~\cite{aouani2011plasmonic} (dashed-lines).
	}
\end{figure}

Here, two different simulations are separately done one for Alexa Fluor 647 and one for Rhodamine 6G molecules as the fluorescent molecules, while the output emission is detected at two perpendicular arc-ports on the hemisphere encircling the disk as illustrated in Fig.~\ref{fig:heykel}. The long-time average, as well as the ensemble average of the intensity are presented in Fig.~\ref{fig:heykel_emi}. The results are in a good agreement with those reported in~\cite{aouani2011plasmonic} (see Fig.~\ref{fig:heykel_emi}, dashed lines), i.e., for Alexa Fluor 647 the peak intensity is observed at a polar angle of around 27 degrees while for Rhodamine 6G the emission become concentrated at the zero polar angle.

\subsection{Fluorescence coherence}
\begin{figure}[!t]
	\centering
	\includegraphics[width=0.99\columnwidth]{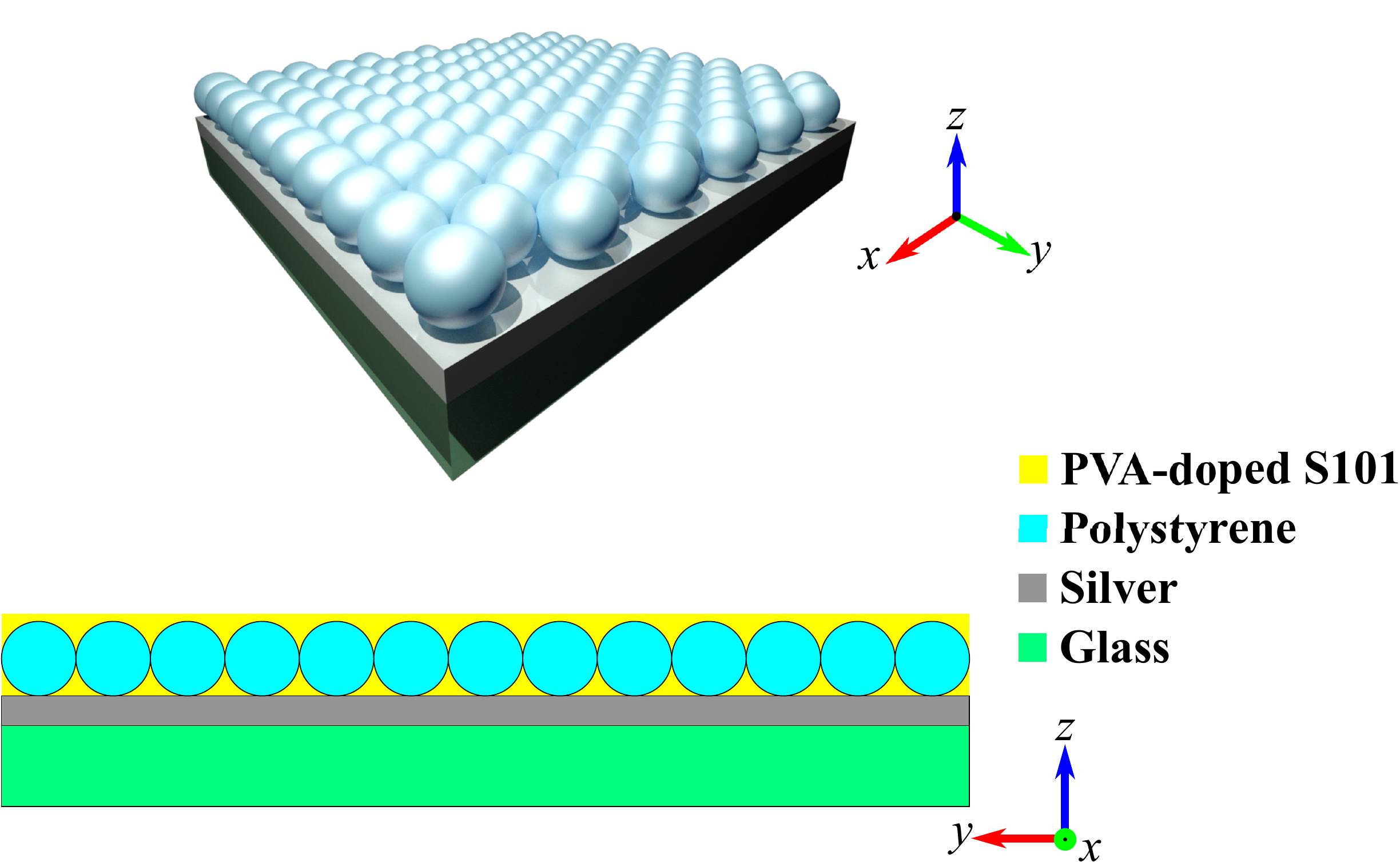}
	\caption{\label{fig:Shi_struct}Schematic of the HPPS proposed by Shi et al.~\cite{shi2014coherent}. The $y$-$z$ cross-sectional view of the structure is shown at the bottom, in which the fluorescent-doped PVA filling is also marked.
	}
\end{figure}

The main aim of developing the proposed method is to capture the statistical attributes of a many-emitter system. In this sense, the method is utilized to simulate the response of a set of fluorescent molecules positioned adjacent to a photonic crystal (PC) constructed by triangular arrangement of polystyrene spheres of 500~nm diameter. 
This PC is placed on top of a 200~nm thick silver slab to form an HPPS as shown in Fig.~\ref{fig:Shi_struct}.
The incident pump ($\lambda=532$~nm) works continuously in $y$-direction while the output emission is detected in $z$-direction.
The capability of this structure in producing a coherent light from fluorescent spontaneous emissions was previously reported by Shi et al.~\cite{shi2014coherent}.
Coherence of a fluorescent light is one of the statistical phenomena, which to the best of authors' knowledge has not yet been numerically addressed.
In this case, the fluorescent material, fluorophore-doped polyvinil alcohol (PVA), forms a 50~nm thick layer on top of the HPPS and also fills the vacancy between spheres (see the cross-sectional view in Fig.~\ref{fig:Shi_struct}). 
In order to keep the numerical test-case substantially similar to the reported experimental setup, properties of the fluorophores, i.e. excitation and emission wavelength are set according to the physical properties of Sulforhodamine 101 (S101) as $\lambda_{ex}=$~575~nm and $\lambda_{em}=$~591~nm, respectively (Table~\ref{tab:flupara}). A $y$-directed 532~nm continuous-wave is used to pump the structure from the center of $x$-$z$~plane.

\begin{figure*}
	\centering
	\includegraphics[width=0.6\textwidth]{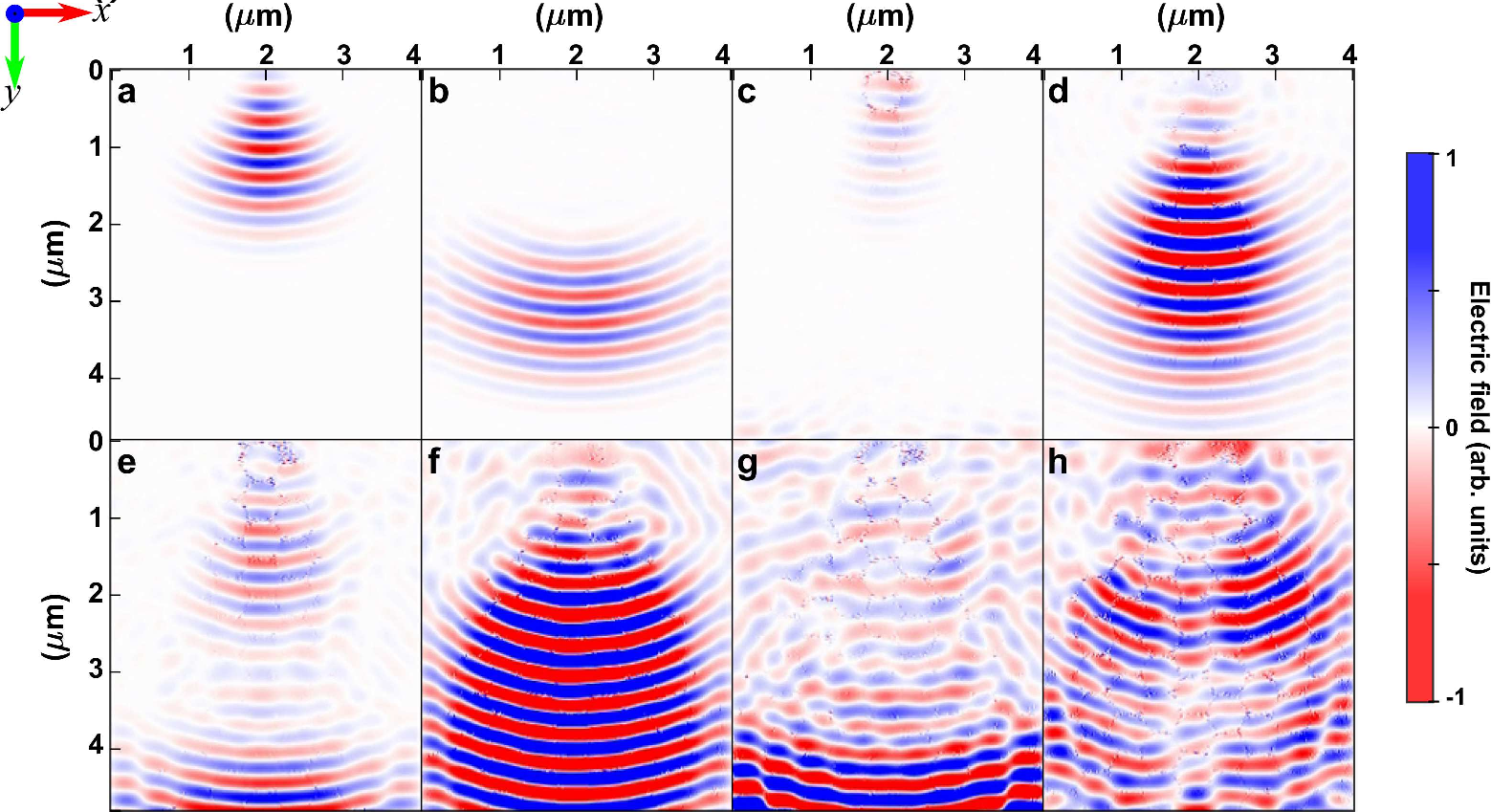}
	\caption{\label{fig:shi_field}The time evolution of the electric field produced by a pulse-train in the photonic-crystal part of the HPPS proposed by Shi et al.~\cite{shi2014coherent}. The field distribution on a surface passing through the center of the polystyrene spheres in the $x$-$y$~plane (see the structure in Fig.~\ref{fig:Shi_struct}) is illustrated at selective time-steps. The $y$-directed incident pump pulse enters the domain of the structure from the center of $x$-$z$~plane ($y=0$). Here, the first four pulses of the train are depicted. The fluorescent emission of the molecules is observed as small spots disturbing the pump field.}
\end{figure*}

In Fig.~\ref{fig:shi_field}, the time-evolution of the field distribution is plotted on a surface passing through the center of the polystyrene spheres in the $x$-$y$~plane. Since the low-intensity fluorescent emissions are masked by the pump intensity that would result in a non-clear field distribution, here, the results are plotted merely for a pulse-train. It is worth noting that for all other simulations of the current test-case, the continuous wave is used. It is observed that the first pump pulse passes through the domain without any fluorescent molecule emission (Fig.~\ref{fig:shi_field}b). The fluorescent emission is seen in Figs.~\ref{fig:shi_field}c-h since the molecules have gained the excitation energy required for emission. The honey-comb like pattern (seen more clearly in Fig.~\ref{fig:shi_field}g and h) is caused by the fluorescent emission of molecules filling the vacancy between the polystyrene spheres. 

\begin{figure}[!t]
	\centering
	\includegraphics[width=0.87\columnwidth]{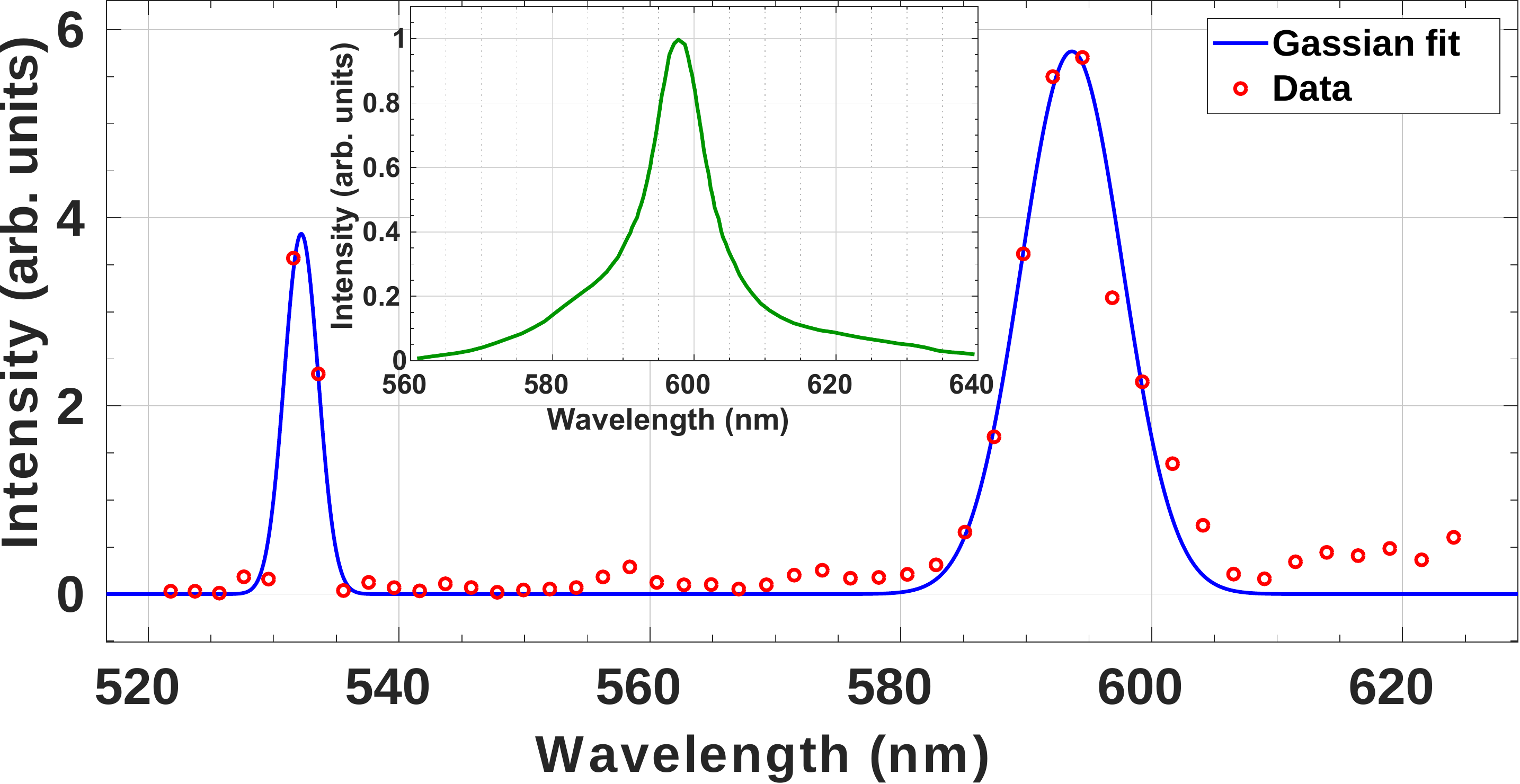}
	\caption{\label{fig:Shi}Wavelength spectrum that is obtained using the proposed numerical method. The peak corresponds to the reflected portion of the excitation wave is observed at $\lambda=532$~nm and the emission peak occurs at $\lambda=594$~nm with the bandwidth of $|\Delta\lambda|=7.1$~nm. The inset shows the emission spectrum as reported by Shi et al.~\cite{shi2014coherent}.
	}
\end{figure}

In order to estimate the degree of temporal coherence, two different approaches has been employed; in one approach, the temporal coherence function (TCF) is utilized, which is the auto correlation of the signal, $$\Gamma(\tau)=\langle u(t+\tau)u^*(t)\rangle,$$ where angle brackets and superscript $*$ denote the time averaging and complex conjugate, respectively~\cite{goodman2015statistical}. Here, $u(t)$ is the electric field. The degree of coherence is conclusively determined as $\gamma(\tau)={\Gamma(\tau)}/{\Gamma(0)}$ and thus, the coherence time is
\begin{equation}
\tau_c=\int_{-\infty}^{\infty}|\gamma(\tau)|^2d\tau.\label{eq:tc}
\end{equation}
Using this approach for the present test-case, the coherence time is $\tau_c = 1.14\times10^{-13}$ sec, this is approximately equivalent to the wavelength bandwidth of $|\Delta\lambda|=6.9$~nm.
In the other approach, the desirable wavelength spectrum is obtained using the Fourier transform of the time-varying electric field. For the present test-case, this spectrum is shown in Fig.\ref{fig:Shi}. By fitting a Gaussian function into the figure, the emission bandwidth of the peak is calculated to be approximately $|\Delta\lambda|=7.1$~nm.

It must be noted that in the present work, a perfectly ordered structure is modeled while any experimental test is subject to fabrication defects. 
These structural defects lead to a wider bandwidth and therefore, a lower temporal coherence can be detected for the experimental setup.
This is due to the fact that any defect/disorder perturbs the Hamiltonian of the structure, which in turn broadens the band of energy~\cite{Mukherjee:12}. 
Considering this issue, there is a good agreement between the result obtained using the proposed numerical method and the $|\Delta\lambda|=9$~nm bandwidth reported in the literature~\cite{shi2014coherent}. 

\begin{figure}[!t]
	\centering
	\includegraphics[width=0.99\columnwidth]{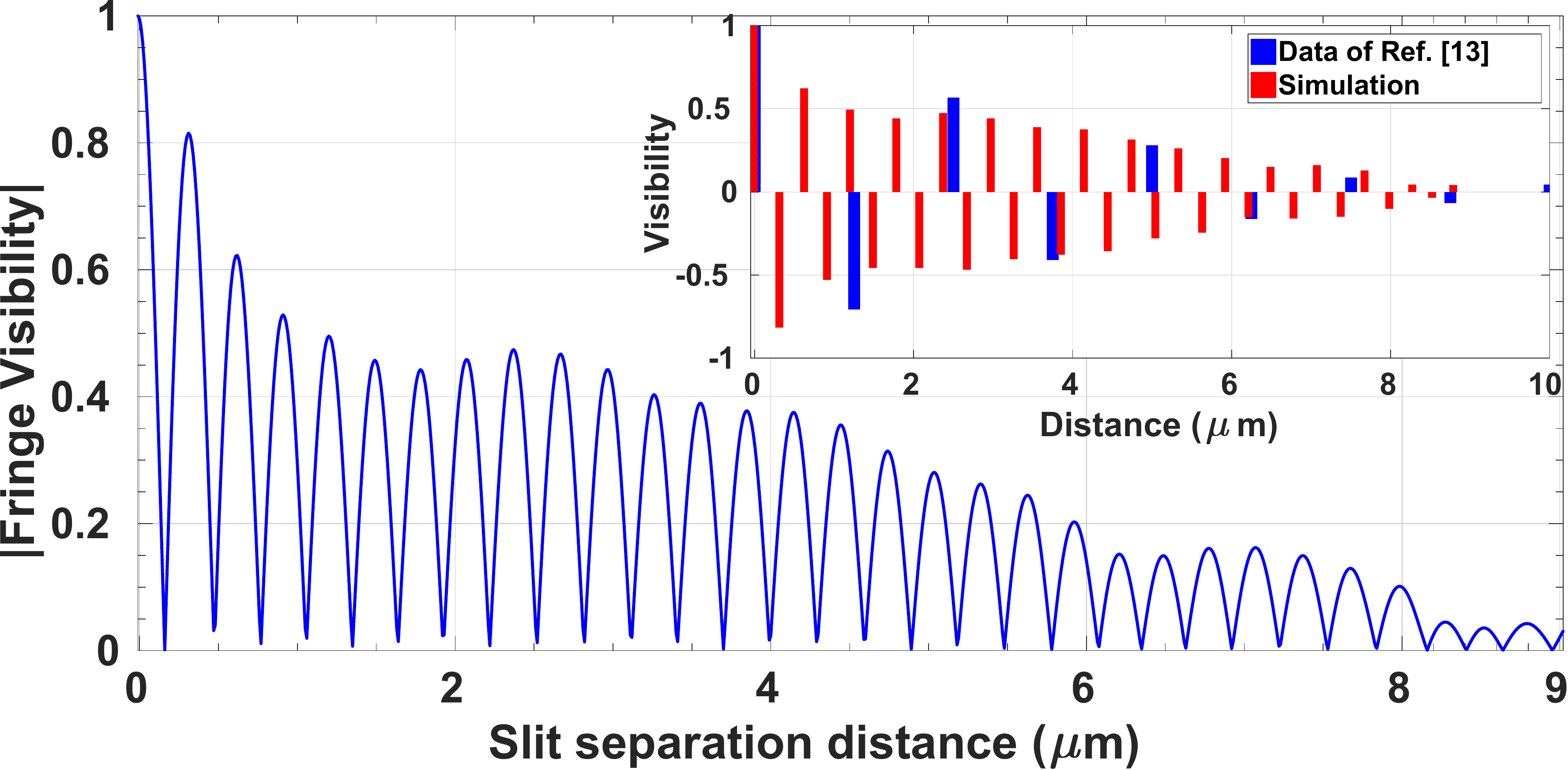}
	\caption{\label{fig:Shi_vis}Young's double-slits fringe visibility as a function of the separation distance of the double-slits calculated using the proposed numerical method. The main plot shows the absolute value of visibility, while in the inset, the maxima and minima of visibility are plotted as red bars. The results reported in Fig.~4c of Ref.~\cite{shi2014coherent} are also plotted (blue bars) for comparison.}
\end{figure}

On the other hand, the most reliable and extremely practical approach to the estimation of spatial coherence is the Young's double slits technique~\cite{martienssen1964coherence} for which, the interference fringe visibility is analytically calculated as a function of the separation distance between slits as~\cite{goodman2015statistical} $$\mathcal{V}_{ij}=\frac{2\sqrt{u_i^2(t)u_j^2(t)}}{u_i^2(t)+u_j^2(t)}M_{ij}.$$ In this equation, $$M_{ij}=\frac{\langle u_i(t)u^*_j(t)\rangle}{\sqrt{\langle u_i^2(t)\rangle\langle u_j^2(t)\rangle}}$$ is the mutual coherence function between electric fields $u_i(t)$ and $u_j(t)$ detected at two distinct points $i$ and $j$ placed on a plane perpendicular to the direction of detection, which represent the positions of the slits. 
For the present test-case, the fringe visibility is plotted in Fig.~\ref{fig:Shi_vis}.
It is evident that visibility significantly decreases as the slits separation distance increases beyond $4\:\mu$m and practically, no interference pattern can be observed for a separation distance of larger than $8\mu$m.
This is in close agreement with the result reported in the reference~\cite{shi2014coherent}; clearly visible interference fringes were observed for the slit-separation distance of $4\mu$m and less clear interference fringes were recognized for the slit-separation distance of $7\mu$m. Nevertheless, it is noteworthy that using numerical simulation, it can be observed (Fig.\ref{fig:Shi_vis}) how the fringe visibility varies for a wide range of separation distances.
	In Fig.~\ref{fig:Shi_vis} (inset), the maxima and minima of the visibility calculated using the proposed numerical method are plotted. In this figure, the intensity of modes as a function of propagation length as reported by Shi et al.~\cite{shi2014coherent} is also illustrated for comparison.


\section{Extending the proposed method}
One of the main advantages of the proposed method is its capability to be modified for other applications beyond many-spontaneous-emitters by modifying the probability functions. 
In this section, the proposed method is modified to capture the statistical behaviour of a many-emitter system, which also exhibits stimulated transitions besides the spontaneous transitions. 
To this end, it is needed to modify the proposed algorithm in order to also include the probability function $P_{10stim}$ required to model the stimulated emission transition. This probability should be based on transition-field coupling and thus can be defined analogous to Eq.~\ref{eq3} as
\begin{equation}
P_{10stim}(t_n)=\exp\left(-{\left({p_{stim}}(t_n)-p_{stim,max}\right) ^2}/{2\sigma_{stim}^2}\right),\label{eq:stimP}
\end{equation}
where $p_{stim}$ is obtained using the following harmonic oscillator equation
\begin{equation}
\ddot{{p}}_{stim}+\gamma_{stim}\dot{{p}}_{stim}+\omega_e^2{p_{stim}}=({e^2}/{m})|\mathbf{E}(t)|.\label{eq:pstim}
\end{equation}
Here, the subscript $stim$ corresponds to the stimulated transition and $\omega_e$ is the frequency of the stimulating photons. It must be noted that upon transition a wave-packet of the form 
\begin{equation}
\begin{split}
\mathbf{F}_{stim}(\tilde{t})=\mathbf{F}_{stim,0}&\exp\left(-\frac{(\tilde{t}-t_{stim,0})^2}{\sigma_{stim,e}^2}\right)\\
\times&\sin(\omega_e(\tilde{t}-t_{stim,0}))
\end{split}
\label{eq:stimPulse}
\end{equation}  
is emitted. Since the polarization of a photon released in stimulated emission must be aligned with the polarization of the stimulating photon, the amplitude of the wave packet is $\mathbf{F}_{stim,0}=F_0\hat{\mathbf{E}}(t)$, where $\hat{\mathbf{E}}(t)$ represents the unit vector aligned with the electric field at the position of the molecule and the instance of stimulated transition.
It is worth noting that the previously proposed semi-classical FDTD approaches (for example see~\cite{Chang:04} and~\cite{PhysRevA.52.3082}) incorporate the statistically averaged quantities from a deterministic viewpoint. 
Therefore, those methods are only capable of modeling a bulk of emitting material, in contrast to many-emitter systems with localized sources that are successfully simulated using the present method. 

\begin{figure}[!t]
	\centering
	\includegraphics[width=0.4\columnwidth]{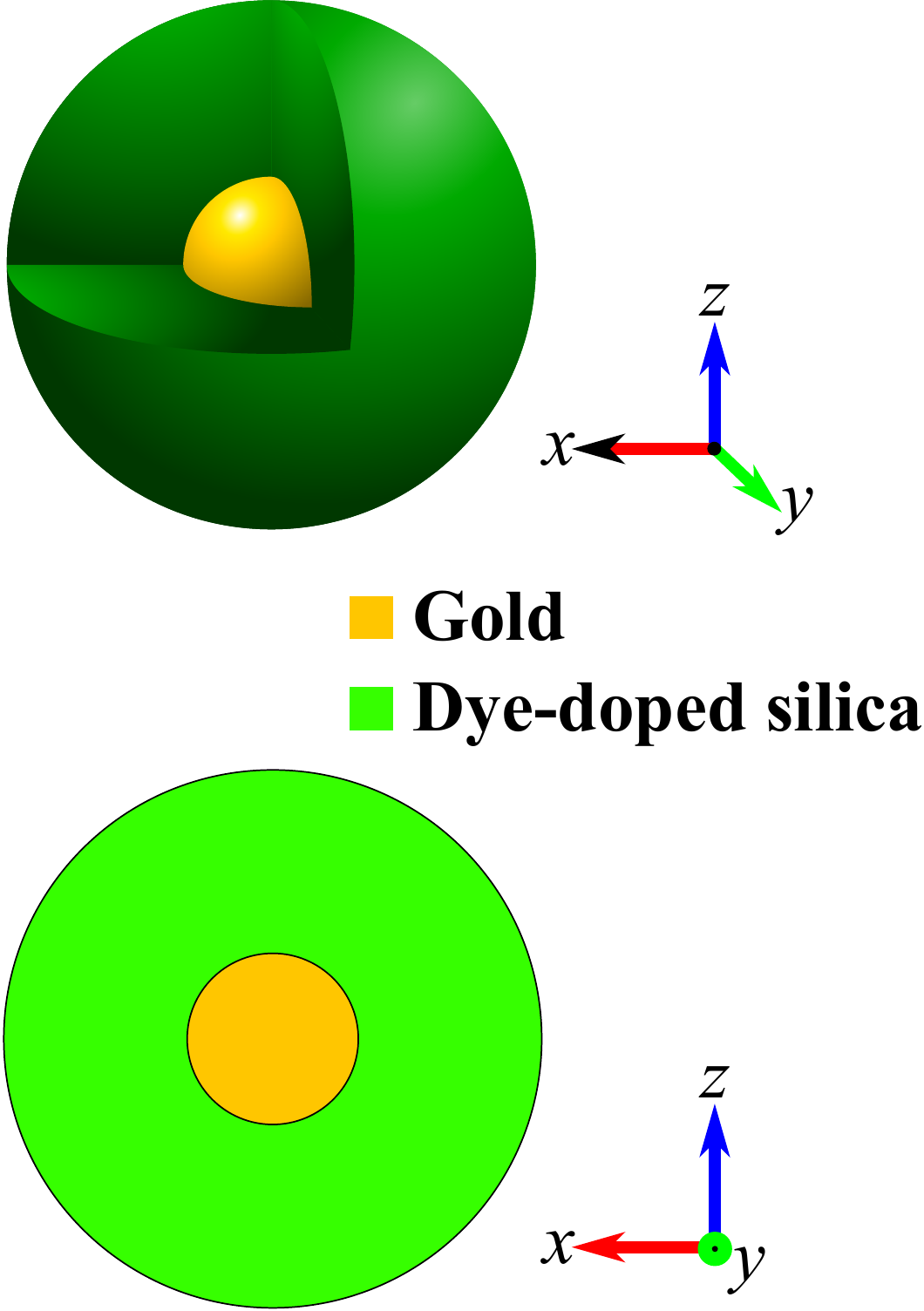}
	\caption{\label{fig:GNP}Schematic of the core-shell type SPASER.}
\end{figure}

\begin{figure}[!t]
	\centering
	\includegraphics[width=0.99\columnwidth]{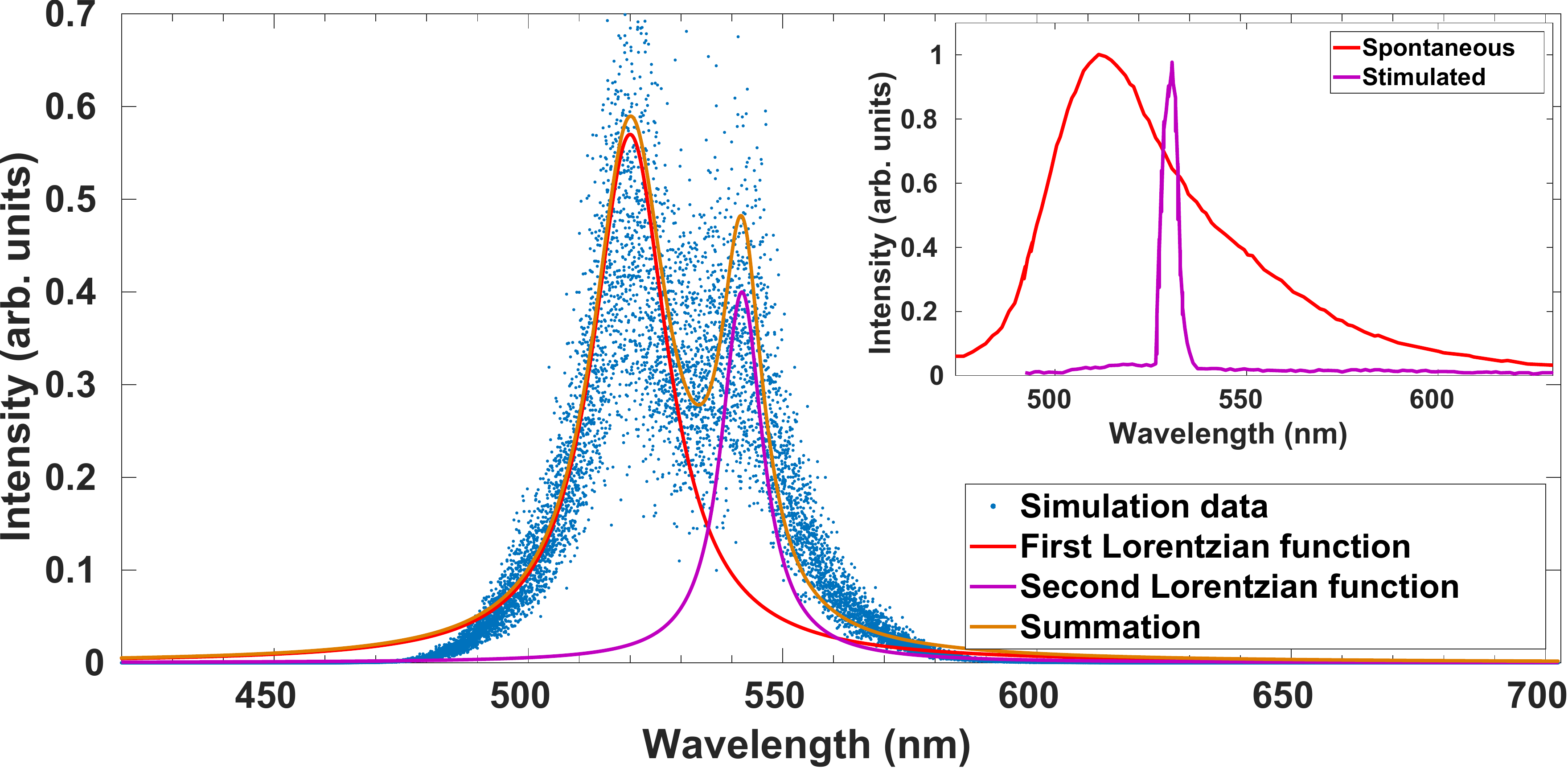}
	\caption{\label{fig:GNP_spec}Emission spectrum of SPASER. The Lorentzian functions and their summation are matched to the data and shown in red, purple, and brown, respectively. The results reported by Noginov et al.~\cite{Noginov2009} are shown in the inset.}
\end{figure}

In order to verify the performance of the proposed method, a core-shell type surface plasmon amplification by stimulated emission of radiation (SPASER) is simulated and the numerical results are compared to the experimental results reported by Noginov et al.~\cite{Noginov2009}. The structure is schematically shown in Fig.~\ref{fig:GNP}; the core is made of gold with the diameter of 14nm, which is enclosed by an Oregon Green 488 (see Table~\ref{tab:flupara}) doped silica shell of 44nm diameter.
Here, the goal is only to investigate the performance of the method in simulating stimulated emission and hence, the pump mechanism is considered to be out of scope.
In this sense, an initial energy-level population inversion is considered for the dye-molecules, i.e. all Oregon Green 488 molecules are initially at level 2.
The emission spectrum is shown in Fig.~\ref{fig:GNP_spec}, which shows two peaks around $\lambda_{spon}=520$~nm corresponding to the spontaneous emission and $\lambda_{stim}=540$~nm corresponding to the stimulated emission. Here, the peak wavelengths are calculated by matching two Lorentzian functions to the data as illustrated in Fig.~\ref{fig:GNP_spec}. The agreement between these results and those reported in~\cite{Noginov2009}, i.e. $\lambda_{spon}=520\pm20$~nm and $\lambda_{stim}=531$~nm, approves the validity of the proposed method for many-spontaneous/stimulated-emitters (see the inset of Fig.~\ref{fig:GNP_spec}). However, the slight deviation in $\lambda_{stim}$ is caused by the so-called stair-case error generally occurs in the FDTD method. In the experimental case, the stimulated emission is resulted from the feedback provided by the surface plasmons of the core inside a cavity of a spherical shape. In the FDTD simulation, on the other hand, the spheres cannot exactly be fitted by the Cartesian grid. Here, mapping of the faces onto the computational cells leads to the appearance of some stairs at the surface of the spheres. Thus, the resonance modes of the numerically modeled simulated sphere are slightly different from those of a real spherical cavity and consequently, the peak of the stimulated emission obtained in the simulation is slightly different than that of the experiment. In order to alleviate this error one needs to either use a highly refined domain discretization that needs prohibitively intense computations or utilize a body-fitted grid, which is out of the scope of the present work. 

\section{Conclusion}
A numerical method was developed, which is capable of capturing the statistical nature of the interactions between a group of emitters and the surrounding EM field.
The proposed algorithm was devised in a manner that is basically consistent with physical principles. 
The method has been validated for three different many-emitter systems; in the first test-case, the directionality of the fluorescent emission is captured. 
In the second test-case, the spatial and temporal coherence are simulated for an ensemble of spontaneous emitters.
This important statistical attribute can be numerically captured merely by utilizing the proposed probabilistic approach.   
In order to show the capability of the proposed method beyond the spontaneous emitters, in the last test-case, the stimulated emission of a SPASER is simulated. 

It is evident that the applications of the proposed method are not limited to the test-cases simulated in this work; the method can be utilized to capture other statistical attributes, e.g. transition rates. In addition, by using different probability functions and/or including two or four energy-levels, the method can be further developed for numerous cases with various types of emitters. Acquiring intensity-dependent probability functions is also a means to obtain more realistic physics. Moreover, nonlinear effects, e.g. two-photon absorption or up-conversion, are possible to be captured by adding virtual states and considering corresponding transitions in the algorithm.

It is worth noting that using the proposed method, the run-times are increased by less than $5 \%$ comparing to a sole electromagnetic solver; however, the computational costs (memory and run-time) depend on the number of emitters.
Considering the complexities and costs associated with experiments, the proposed method is a promising means to facilitate new designs for optical structures and advance the fundamental understanding of the statistical phenomena in the field of light-matter interaction.

%

\end{document}